\documentclass[
superscriptaddress,
twocolumn,
nofootinbib,
 amsmath,amssymb,
 aps,
]{revtex4-2}
\usepackage{xcolor}

\usepackage{graphicx}% Include figure files
\usepackage{dcolumn}% Align table columns on decimal point
\usepackage{bm}% bold math
\usepackage{hyperref}% add hypertext capabilities
\usepackage{comment}

\newcommand{\cevns}{CE$\nu$NS}

\begin{document}

\preprint{APS/123-QED}

\title{First constraints on the coherent elastic scattering of reactor antineutrinos\\ off xenon nuclei}

\newcommand{\Mephi}{\affiliation{National Research Nuclear University MEPhI (Moscow Engineering Physics Institute), Moscow, 115409, Russia}}

\newcommand{\KI}{\affiliation{National Research Center ``Kurchatov Institute'', 1 Akademika Kurchatova sq., Moscow, 123182, Russia}}

\newcommand{\Tomsk}{\affiliation{National Research Tomsk Polytechnic University, 30 Lenin ave, Tomsk, 634050, Russia}}

\newcommand{\JINR}{\affiliation{Joint Institute for Nuclear Research, 6 Joliot-Curie St, Dubna, Moscow region 141980, Russia}}

\newcommand{\FIAN}{\affiliation{P.N. Lebedev Physical Institute of the Russian Academy of Sciences, 53 Leninskiy Prospekt, Moscow, 119991, Russia}}

\newcommand{\INR}{\affiliation{Institute for Nuclear Research, 7a 60-letiya Oktyabrya ave, Moscow, 117312, Russia}}

\newcommand{\RTU}{\affiliation{MIREA - Russian Technological University, Lomonosov Institute of Fine Chemical Technologies, 86 Vernadsky Avenue, Moscow, 119571, Russia}}

\newcommand{\UN}{\affiliation{University of Naples Federico II, Corso Umberto I 40, Naples, 80138, Italy}}

\author{D.Yu.~Akimov}\email[Contact author:]{DYAkimov@mephi.ru} \Mephi 

\author{I.S.~Alexandrov}\Mephi\Tomsk

\author{V.A.~Belov }\KI\Mephi

\author{A.I.~Bolozdynya}\email[Contact author:]{AIBolozdynya@mephi.ru} \Mephi 

\author{A.V.~Etenko}\KI \Mephi

\author{A.V.~Galavanov}\Mephi

\author{Yu.V.~Gusakov}\JINR

\author{A.V.~Khromov}\Mephi
 
\author{A.M.~Konovalov}\Mephi\FIAN

\author{V.N.~Kornoukhov}\Mephi\INR

\author{A.G.~Kovalenko}\KI\Mephi

\author{E.S.~Kozlova}\Mephi

\author{A.V.~Kumpan}\Mephi

\author{A.V.~Lukyashin}\Mephi\RTU

\author{A.V.~Pinchuk}\Mephi

\author{O.E.~Razuvaeva}\email[Contact author:]{or.firefox@gmail.com} \Mephi

\author{D.G.~Rudik}\Mephi\UN

\author{A.V.~Shakirov}\Mephi

\author{G.E.~Simakov}\Mephi\KI

\author{V.V.~Sosnovstsev}\Mephi

\author{A.A.~Vasin}\Mephi

\collaboration{RED-100 collaboration} \noaffiliation

%\date{\today}

\begin{abstract}
RED-100 is a two-phase emission detector with an active volume containing 126~kg of liquid xenon. The detector was exposed to the antineutrino flux of about $1.4 \cdot 10^{13}~$cm$^{-2}$s$^{-1}$ at a distance of 19~m from the 3.1~GW Kalinin Nuclear Power Plant (KNPP) reactor core. The comparison of data from 331~kg$\cdot$days with the reactor on and 106~kg$\cdot$days with the reactor off shows no statistically significant excess and allows to put constraints on coherent elastic interactions of antineutrinos with xenon nuclei.
\end{abstract}

\maketitle

%\tableofcontents

\section{\label{sec:intro}Introduction}

Coherent elastic neutrino-nucleus scattering (\cevns{}) is a Standard model process mediated by the neutral current of the weak interaction~\cite{Freedman, Kopeliovich:1974mv}. The cross-section of this process dominates among other interactions of low energy neutrinos (E$_{\nu}\lesssim50$~MeV) with matter due to interference of neutrino-nucleon scattering amplitudes. The \cevns{} cross-section~\cite{Lindner:2016wff} can be written as
\begin{equation} \label{eq1}
      \frac{d\sigma}{dT} \simeq \frac{G_F^2Q_w^2M}{4\pi}\Big(1-\frac{T}{T_{max}}\Bigr)F^2_{nucl}(q^2), 
\end{equation} 
where $G_F$ is the Fermi constant, $M$ is the mass of the nucleus and $T$ is energy of a nuclear recoil. The weak charge of a nucleus is
\begin{equation} \label{eq2}
    Q_{w} = N-(1-4\sin^2\theta_W)Z, 
\end{equation} 
where $Z$ and $N$ are the numbers of protons and neutrons in a target nucleus, while $\sin\theta_W$ is the electroweak mixing angle~\cite{Erler:2004in}. The cross-section vanishes for the backward neutrino scattering corresponding to the maximal energy of a nuclear recoil
\begin{equation} \label{eq3}
    T_{max}=2E_\nu^2/(M+2E_\nu),
\end{equation}
where $E_\nu$ is the incident neutrino energy.

The degree of interference of individual neutrino-nucleon scattering amplitudes is characterized by the value of a nuclear form factor $F^2_{nucl}$ depending on the value of momentum transfer squared $q^2$. This form factor reflects a spatial distribution of nucleons within a nucleus relative to the transferred momentum wavelength~\cite{Helm:1956zz,Klein:1999qj}.

The delay of more than forty years between the \cevns{} prediction and its first observation~\cite{COHERENT:2017ipa} illustrates the challenge of finding a combination of a very bright neutrino source and an extremely sensitive detector. While the suitable energy spectrum and flux magnitude are required from the former, the latter should provide a sufficient target mass and a low energy threshold. The field of \cevns{} research includes experiments at pion decay at rest sources ($\pi$DAR)~\cite{Akimov:2022oyb, Shoemaker:2021hvm, Baxter:2019mcx}, nuclear reactors~\cite{CONUS:2020skt, Ackermann:2024kxo, nGeN:2022uje, Colaresi:2022obx, CONNIE:2021ggh, CONNIE:2024pwt, RED-100:2019rpf, Ricochet:2022pzj, Ricochet:2023nvt, TEXONO:2018fse, Karmakar:2024ydi, Nucleus:2022ijh, NEON:2022hbk, Chaudhuri:2022pqk, Yang:2024exl, Alfonso-Pita:2023frp} and underground dark matter search experiments able to probe solar neutrinos~\cite{OHare:2021utq, LZ:2022lsv, XENON:2023cxc, PandaX:2022aac}. At the moment of writing, three \cevns{} detection results are reported by the COHERENT collaboration~\cite{COHERENT:2021yvp} at the Spallation Neutron Source (CsI~\cite{COHERENT:2017ipa,COHERENT:2021xmm}, Ar~\cite{COHERENT:2020iec}, Ge~\cite{Adamski:2024yqt}) and two from the dark matter search experiments with ton-scale two-phase xenon detectors~\cite{PandaX:2024muv, XENON:2024ijk}.

The experiments at nuclear reactors promise both valuable scientific and technological results. The former include constraints on non-standard neutrino interactions~\cite{Barranco:2005yy, Giunti:2014ixa} (particularly induced by low-mass mediators), while the latter are associated with the \cevns{} detectors potential for nuclear nonproliferation~\cite{Barbeau:2002fg, Hagmann:2004uv, Bernstein:2019hix}. The only \cevns{} measurement at reactors to date is reported by the Dresden-II experiment~\cite{Colaresi:2022obx} using a germanium detector. A tension between this claim and the recent constraint from CONUS~\cite{Ackermann:2024kxo} is to be resolved by further measurements.

RED-100~\cite{Akimov:2022xvr} is a two-phase xenon detector~\cite{PismaZhETF.11.513, EmDet1994} with the largest sensitive mass and the heaviest target nucleus out of all reactor \cevns{} experiments to date. Such a combination together with the sensitivity and scalability of the two-phase technique is promising for scientific research as well as reactor monitoring applications.
Recently another experiment (RELICS) with a similar concept was proposed for \cevns{} observation at the Sanmen Nuclear Power Plant~\cite{Cai:2024bpv}.
This work is devoted to the first constraints on \cevns{} from the RED-100 exposition at the Kalinin Nuclear Power Plant~(KNPP)~\cite{physics5020034}.

\section{\label{sec:RED100}The RED-100 experiment}

\subsection{Experimental setup}

\begin{figure}[htb]
\includegraphics[width=1\linewidth]{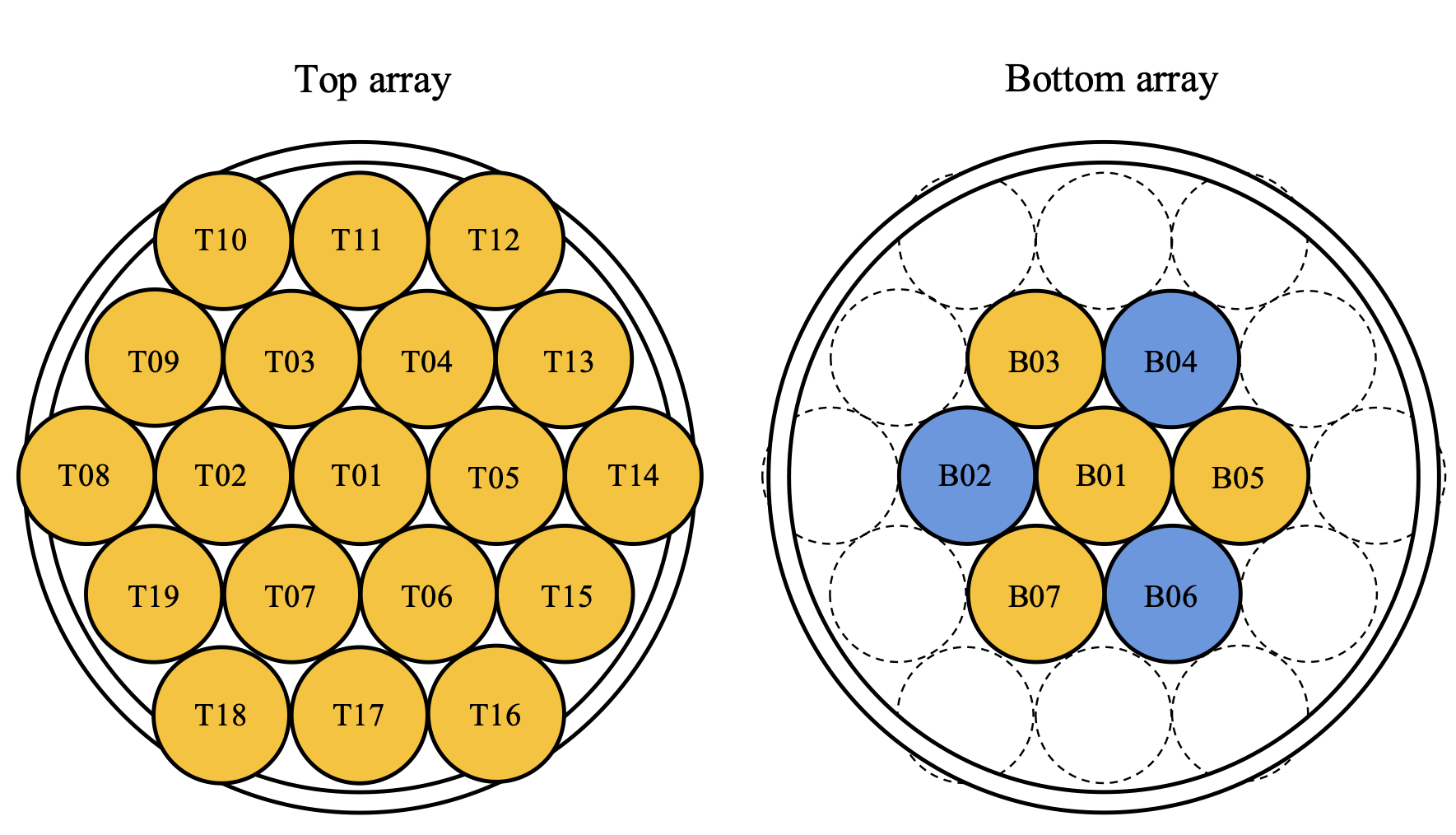}%
\caption{\label{fig:PMTs} The top-view scheme of the PMT arrays of RED-100. PMTs B02, B04, and B06 (blue color) were operated at lower voltage (see text).}
\end{figure}

The cylindrical sensitive volume of the RED-100 detector has a diameter of 36~cm and a height of 41.5~cm and contains about 130~kg of liquid xenon~(LXe). Ionizing radiation produces both excitation and ionization of xenon atoms in the sensitive volume. The excitation leads to the scintillation flash coincident with the moment of interaction. The ionization electrons drift to the surface of the liquid and are extracted into the 0.9~cm thick gas gap filled with the xenon vapor at about 1.3~atm, where the electric field is strong enough to cause electroluminescence.
It is generated all along the paths of electrons from the LXe surface to the anode electrode.  The light is collected by a PMT array of 19 Hamamatsu R11410-20 units~\cite{Akimov:2015aoa, Akimov:2016fjf} located above the anode. 
Seven PMTs of the bottom array facilitate the detection of primary scintillation.
The scheme of PMTs positions is shown in fig.~\ref{fig:PMTs}. The drift field strength in the largest part of the liquid volume but the upper 1.0~cm
is about 218~V/cm corresponding to the maximal drift time of 265~$\mu$s to the gas gap. The electric field strength in the 1.0~cm of the liquid above the gate mesh electrode is $2.68\pm0.04$~kV/cm~\cite{CalibrationRED100}, and $4.96\pm0.07$~kV/cm in the gas gap.
The latter accounts for the 2~$\mu$s duration of electroluminescence for a point-like charge.

The RED-100 setup was deployed at Unit 4 of KNPP in 2021~\cite{Akimov:2022xvr,physics5020034}. The detector was exposed to the antineutrino flux of about $1.4 \cdot 10^{13}~$cm$^{-2}$s$^{-1}$
\footnote{Calculated for the specific heat deposition of 205.3~MeV/fission and 6.75 $\bar{\nu}$/fission from ref.~\cite{KopeikinSP1}.}
at a distance of 19~m under the center of the $\sim$3.1~GW thermal power reactor core. The reactor building and construction materials of the unit provide about 50 m.w.e. overburden in the vertical direction and overall reduction of the muon flux by a factor of $\sim$7~\cite{Akimov:2023lsz, Alekseev_2016}. The detector was surrounded by the 5‑cm-thick copper layer and submerged into the water tank, providing about 70~cm of water in any direction~\cite{Akimov:2021hci}.

\subsection{\label{sec:collection}Data collection}

\cevns{} events produce very low energy depositions. For reactor antineutrinos and xenon targets, it is mostly below 1~keV, which results in only several ionization electrons. 
A single electron (SE) signal in RED-100 manifests itself as 20--30 single photoelectron (SPE) signals detected by PMTs and distributed almost uniformly across the electroluminescence duration~\cite{CalibrationRED100}. 
Thus a \cevns{} event consists of dozens to hundreds of SPE pulses distributed over a significant time span of 2 to 5~$\mu$s across 19~PMTs channels. These pulses are small with about 8~mV amplitude (after tenfold external amplification) and 20~ns duration.
To effectively detect such signals a dedicated data acquisition system (DAQ) trigger based on pulse counting was developed~\cite{Naumov:2014sva}.
Each PMT signal is fed to individual discriminators CAEN V895 with a threshold low enough to detect SPE pulses. 
The average measured SPE detection efficiency is about 60\%. 
The signals from the discriminators are sent to the complex digital triggering circuit built using the CAEN V1495 module. 
The circuit counts SPE pulses from the top PMT array in a running 2~$\mu$s window which corresponds to the characteristic electroluminescence duration. 
The \cevns{}-like trigger detection threshold is set to 57 counts, i.e. about 3~ionization electrons to provide the lowest possible energy threshold while keeping the total trigger rate below the 20~Hz maximum allowed by DAQ.

As was shown by RED-100~\cite{Akimov:2016rbs,Akimov:2023xsi} and several other groups~\cite{LUX_2020, XENON_2022}, the big energy deposition from the passing of cosmic muons through the detector is a strong source of the background single electron (SE) like signals. This is important for the RED-100 detector operated with a low overburden.
The electronic shutter was added to the RED-100 electrode structure in order to suppress this effect. It prevents ionization electrons from extraction to the gas gap after the muon passage through the detector by reversing the electric field direction~(see details in~\cite{RED-100:2019rpf}).
The muons are identified by their extremely large scintillation. The signals from the bottom array PMTs (B01, B03, B05, B07 in figure~\ref{fig:PMTs}) are sent to the high-threshold ($\sim$1~V) discriminator and then to the majority two-of-four logic.
The shutter blocking duration varies in the range from~0.3~to~6~ms depending on the muon energy estimated using the width of a scintillation pulse measured by sum signal from PMTs~(B02, B04, B06) operated at a lower voltage.
The trigger is vetoed for a full shutter duration plus 10~$\mu$s. Also, the trigger is vetoed for 300~$\mu$s after gamma events detected by a sum of signals from the bottom PMT array.
As an enhanced measure to suppress time periods with high SE emission rates in the detector (noisy periods), a dedicated veto looking for high SPE rate periods was developed. It counts SPE pulses from the top PMT array in the 50~$\mu$s period and vetoes the trigger if more than 50 pulses are observed. This threshold is chosen to minimize the rate of triggers connected to random coincidences of spontaneous SE signals down to the DAQ recording rate of 20~Hz.

To handle \cevns{}-like events effectively, fast electronics is used, and detailed waveforms with a sampling period of 2~ns are recorded for all PMTs by DAQ for further processing and analysis~\cite{Akimov:2017}. 
The xenon scintillation light yield for nuclear recoils in the reactor \cevns{} region of interest is quite low ($\sim$3 photons for a 1 keV recoil~\cite{szydagis2023review}) making the detection and identification of such a signal unlikely.
We work in S2-only mode which means only the secondary electroluminescent signal originating from the ionization electrons is considered in further analysis.
Hence, it is unnecessary to record data for the maximal ionization drift time of 265~$\mu$s.
For \cevns{} data the recorded waveform duration is reduced to 30~$\mu$s. This choice is made to increase the data readout rate by the cost of degrading the ability to suppress a background from multiple scattering of gamma rays and neutrons. 
The trigger location within the recorded waveform provides about 18~$\mu$s before the candidate to examine the isolation of a signal of interest.
We implemented direct measurement of livetime during the acquisition by counting pulses from a 1~MHz pulser vetoed in the same way as trigger. The average livetime over elapsed real time ratio with \cevns{} trigger is about 60\%.
Accumulated statistics for data collected with \cevns{} trigger during reactor OFF period is 2.5M events for the 0.84 days livetime, while during reactor ON period it is 10.5M events and livetime of 2.63 days.

\section{\label{sec:simulation} \cevns{} signal prediction}
Calculation of sensitivity of the RED-100 detector and evaluation of experimental limit require detailed simulation of \cevns{} signals in the detector. 
Such a simulation uses a proper antineutrino energy distribution to obtain a nuclear recoil spectrum of interest. 
It should be converted then into a spectrum in the units of ionization electrons generated in xenon, which in turn can be recalculated to an observable light signal. 
In this section, we describe each of these steps.

\subsection{\label{subsec:neutrino_simulation}Antineutrino energy spectrum}

The importance of the reactor antineutrino energy spectrum for the \cevns{} signal simulation is related to the challenge of low energy nuclear recoils detection. The part of the antineutrino flux with ${E_{\nu}>8}$~MeV, even low in intensity, can result in nuclear recoils above the detector threshold facilitating \cevns{} observation. 
Several models of reactor antineutrino energy distribution 
can be found in literature~\cite{MullerSP, HuberSP, KopeikinSP1, KurchatovInst1, KurchatovInst2, SM2018, SM2018Table, RENO, RENO_Table, DayaBaySP1, DayaBaySP2HTR, DoubleChoozSP}. In this work we consider spectra suggested by authors from Kurchatov Institute~(KI)~\cite{KurchatovInst1,KurchatovInst2}, one of summation models~(SM2018)~\cite{SM2018, SM2018Table}, results from Daya~Bay~(DB)~\cite{DayaBaySP1,DayaBaySP2HTR} and recent evaluation by authors from Institute for Nuclear Research~(INR)~\cite{DoubleChoozSP} verified using the Double Chooz data~\cite{DoubleChooz:2019qbj}. The first (KI) was used in the calculation of a \cevns{} count rate for the initial RED-100 sensitivity study~\cite{RED-100:2019rpf} and lacks antineutrinos with energy above 8~MeV. The second (SM2018) represents the results of a summation approach. It includes the high energy part of the antineutrino spectrum and allows recalculation for the arbitrary fuel composition. We also compare these two models with the deconvolved antineutrino spectra of DB and INR, both including parts with $E_{\nu}>$8~MeV, however different in intensity.

The calculations performed based on all of these models assume isotropic antineutrino flux produced by a 3.09~GW thermal power VVER-1000 reactor.
All spectra except DB
\footnote{The original works~\cite{DayaBaySP1,DayaBaySP2HTR} provide a single antineutrino spectrum for the fractions of 56.4\%~($^{235}$U), 7.6\%~($^{238}$U), 30.4\%~($^{239}$Pu) and 5.6\%~($^{241}$Pu). As there are no spectra for each of the main fissile isotopes, the recalculation for an arbitrary fuel composition is not possible.}
are recalculated for the main fissile isotopes fractions of 71.7\%~($^{235}$U), 6.8\%~($^{238}$U), 18.4\%~($^{239}$Pu) and 3.1\%~($^{241}$Pu) provided by KNPP. These fractions were considered unchanged throughout the 3 weeks data taking period with the active reactor. The average value of energy released per fission used in these calculations is 204.0~MeV~\cite{MEV,MEV205}. The spectra-averaged differential cross-section for each model was used to obtain the nuclear recoil spectra for the RED-100 as a differential count rate.

\begin{figure}[t]
\includegraphics[width=1\linewidth]{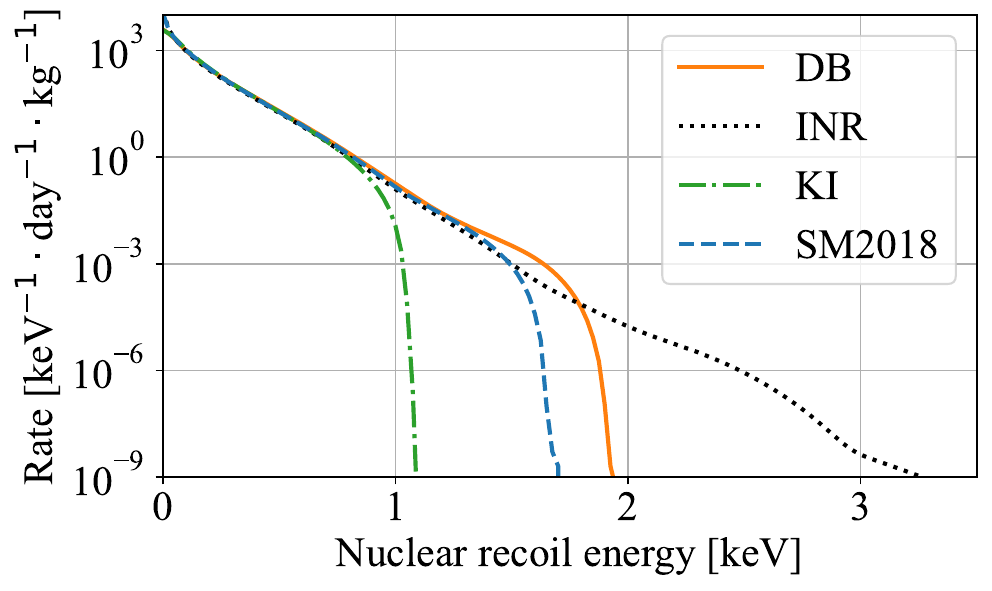}%
\caption{\label{fig_ReactorSP} Energy spectra of \cevns{} xenon recoils for different models of reactor antineutrino energy distributions (see text).}
\end{figure}

The \cevns{} nuclear recoil spectra calculated for each of the antineutrino energy distribution models are shown in figure~\ref{fig_ReactorSP}. 
These nuclear recoil spectra are used as input for the simulation along with the properties of liquid xenon as a sensitive medium and the detector characteristics.

\subsection{\label{subsec:signal_simulation}\cevns{} signal in RED-100}

The simulation of liquid xenon response to nuclear recoils is performed based on the NEST v2.4.0 model~\cite{szydagis2023review} with an input temperature of 169~K, pressure of 1.29~bar and 218~V/cm electric drift field strength. 
The current NEST nuclear recoil model allows the simulation of ionization signals down to 0.2~keV (including the low-energy phenomena reported in~\cite{lenardo2019measurement}) covering most of the analytically calculated recoil spectra. The output of the NEST-based simulation is a number of ionization electrons generated by a nuclear recoil of a given energy at the site of interaction.

\begin{figure}[hbt]
\includegraphics[width=1\linewidth]{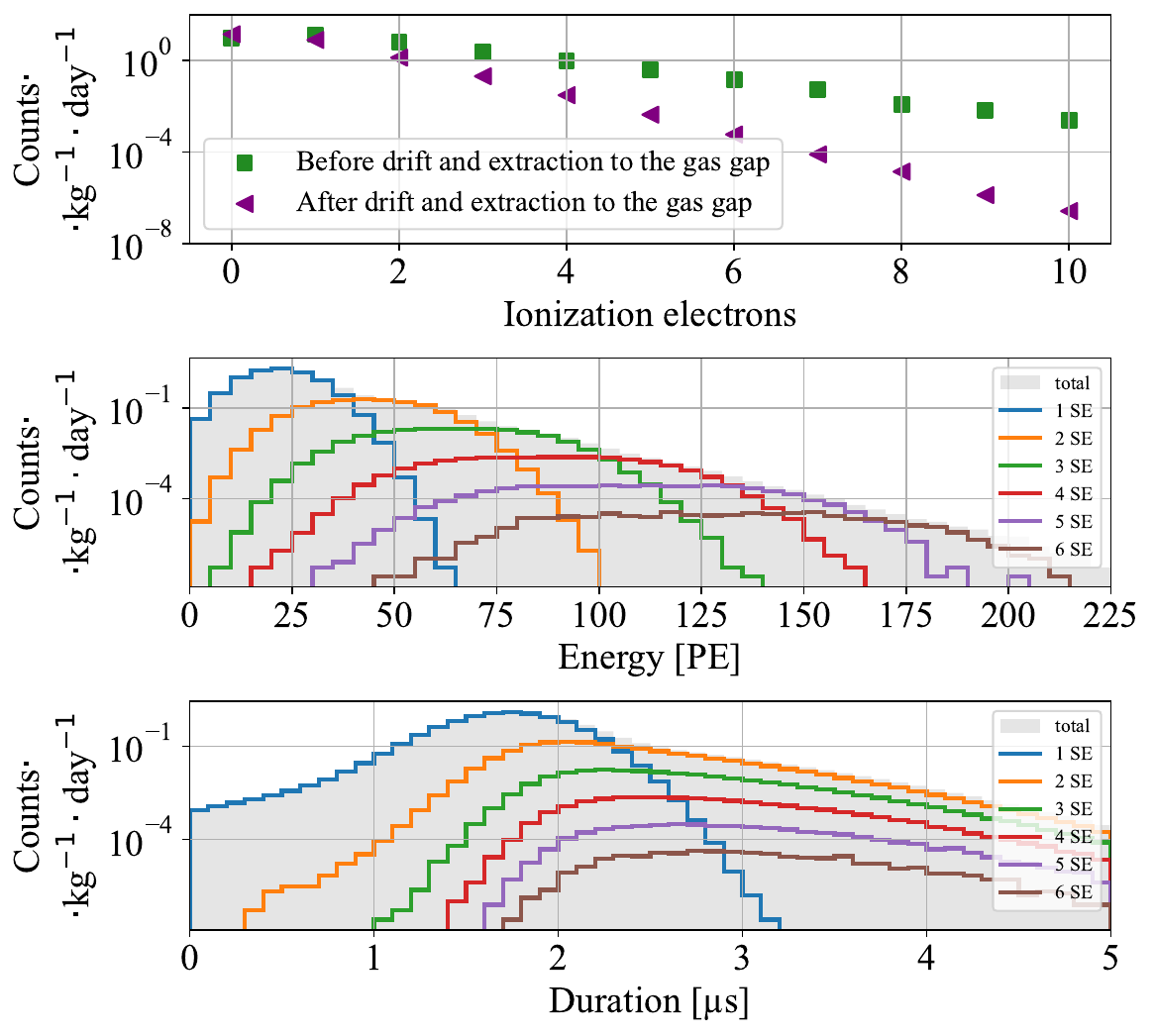}% 
\caption{\label{fig:spectrum_electrons} \textbf{Top:} Simulated \cevns{} spectrum in units of ionization electrons before (green squares) and after (purple triangles) extraction. \textbf{Middle:} The resulting simulated energy spectrum of the predicted \cevns{} signal in units of detected~PE. \textbf{Bottom:} The resulting duration distribution of the predicted \cevns{} signal. All of the plots assume SM2018 and NEST charge yield fluctuations.}
\end{figure}

Before the electroluminescence, a cloud of ionization electrons undergoes two types of losses.
The first is a capture of drifting electrons by electronegative impurities. The characteristic scale of this process is the lifetime of a free electron in liquid xenon. It was measured to be 874$\pm$17~$\mu$s \textit{in situ} based on the signals from cosmic muons~\cite{Akimov:2022xvr, CalibrationRED100}.
The second is the loss of electrons at the liquid-gas interface. It is characterized by the electron extraction efficiency (EEE) coefficient of 32.8$\pm$2.8\% evaluated based on the gamma-calibration data described in~\cite{CalibrationRED100}.
The simulated \cevns{} recoil spectra in units of ionization electrons prior to and after these losses are shown in figure~\ref{fig:spectrum_electrons} (top).
A drift of electrons in liquid xenon is simulated taking into account diffusion of the electron cloud in accordance with the formula described in~\cite{EXO-200:2016qyl} and~\cite{Njoya:2019ldm}. This part of the simulation is crucial for the correct prediction of electroluminescence duration of \cevns{} events.

The S2 part of every signal consists of several ionization electrons and can be represented as a sum of single electron (SE) signals.
The size and duration of the electroluminescence signal produced by a few-electron ionization are simulated using the measured SE parameters and the light distribution over the PMT array which was calculated using light response functions (LRFs). 
The evaluation of SE parameters and LRFs is described in detail in~\cite{CalibrationRED100}.

Events produced by this multi-step simulation consist of a set of photoelectrons (PE) each having detection time and involved PMT.
The resulting energy and duration distributions with the illustrated contribution of each amount of ionization electrons are shown in figure~\ref{fig:spectrum_electrons}.

\section{\label{sec:processing}Data processing}

The 30~$\mu$s-long waveforms of \cevns{} candidate events acquired at KNPP are processed with the help of the REDOffline software~\cite{CalibrationRED100}. 
This software performs pulse finding and parameterization for further analysis.
Only pulses from the top array PMTs with areas larger than the threshold are considered. The threshold is defined as two standard deviations below the mean of a gaussian fit of the SPE area distribution for each PMT.
Sequences of 5 or more pulses, such that there is no more than 500~ns between each two consequent, are identified as a cluster.
A cluster is considered to represent low-energy electroluminescence from an interaction of ionizing radiation with xenon, although it can be associated with the overlap of a few spontaneous SE signals and/or SPE from the scintillation and dark current of PMTs.

The light collection efficiency depends on the cluster position in the horizontal (XY) plane and decreases with radius.
We perform reconstruction of the position and correction of S2 area with a help of LRFs based on calibration data (see~\cite{CalibrationRED100} for details).
In what follows units of detected energy deposition, either PE or ionization electrons, refer to the corrected quantities. 
The ionization electron number is evaluated from a corrected PE number using the measured electroluminescence gain of 27.0~PE per electron~\cite{CalibrationRED100}.

Prior to further discussion of the analysis cuts we define the \cevns{} region of interest (ROI) in the following dimensions: duration, corrected energy, and reconstructed spatial position in the horizontal plane. Cutting on the vertical position of an energy deposition is not possible since S1 is too small to be detected and hence drift distance is unknown for discussed events. We discard events with a duration of less than 1.7~$\mu$s since those are identified as technical background originating from the detector's edge rim. Also, we do not consider events with a duration of more than 4~$\mu$s due to negligible expected \cevns{} rate in this region. It should be mentioned here that the duration we use is calculated from the first SPE to the last one and hence it is bigger than the S2 duration calculated as the full width at half maximum. We also limit cluster energy to be more than 4 and below 7 electrons (110 and 189 corrected PE correspondingly). We only consider clusters 
within a reconstructed radius of 140~mm. This restriction
is introduced due to a decrease in reconstruction precision due to a decline in light collection efficiency near the edge of the detector.

\begin{figure}[hbt]
\includegraphics[width=1\linewidth]{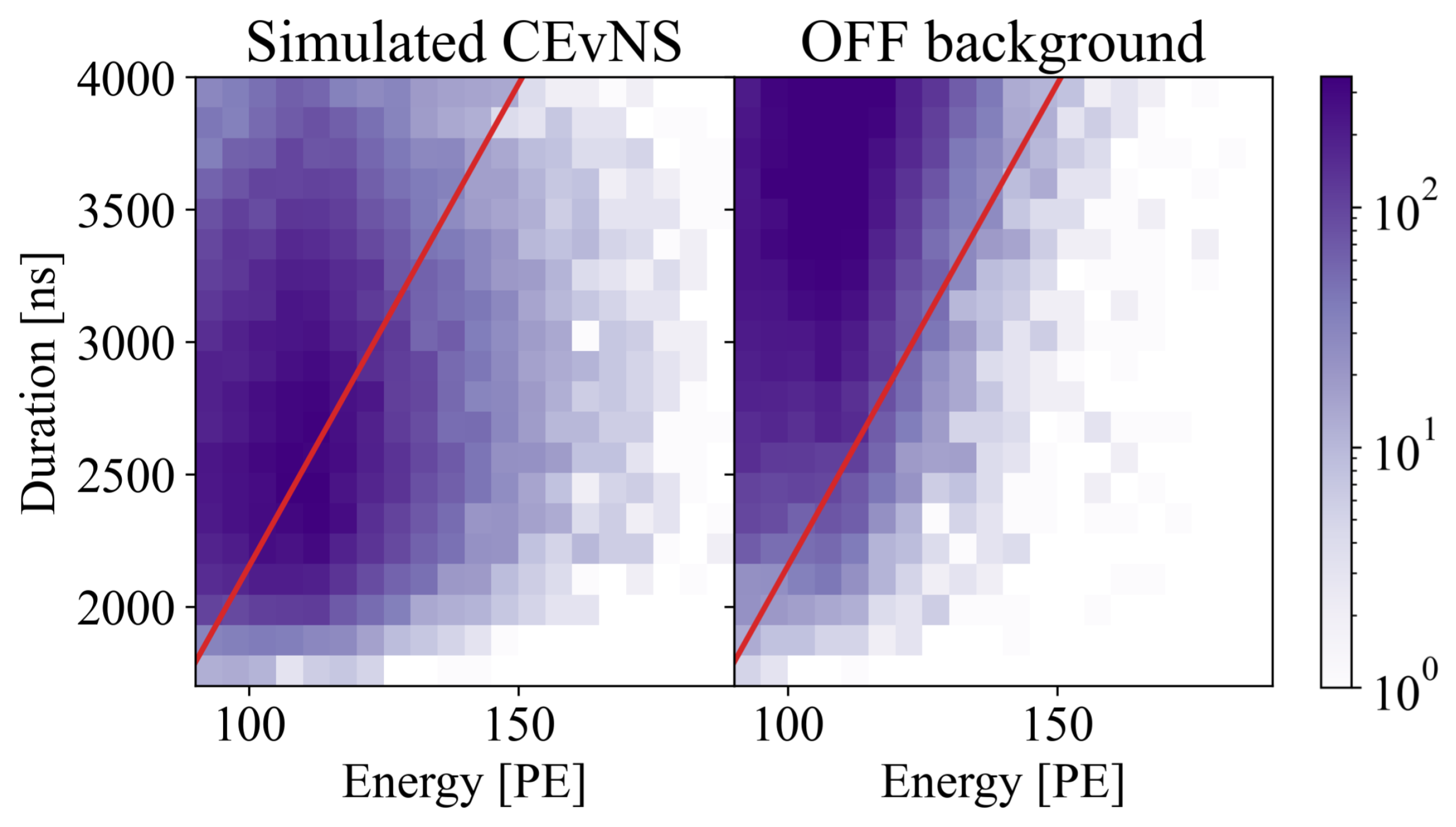}% Here is how to import EPS art
\caption{\label{fig:durationcut} Dependence of the duration cut on the energy. The events above the red line are rejected. Bin values for the simulated \cevns{} events are normalized to the reactor OFF time. Bin values for the simulated \cevns{} events are normalized to the OFF dataset size.}
\end{figure}

We introduce additional limitations of the ROI based on the check of the count rate stability, as a significant correlation of the rate with the ambient temperature fluctuations is observed at KNPP. The data analysis suggests that this correlation is caused by the dependence of discriminators' offsets on temperature leading to changes of SPE detection efficiency by the trigger. 
To cease this effect we exclude a part of the parameter space with a lower number of detected photoelectrons and larger durations. 
In particular, the difference between ON and OFF trigger efficiencies is required to be under 1\% in comparison with OFF efficiency.
The shape of this restriction is shown in figure~\ref{fig:durationcut}. 
This shape ensures that the trigger efficiency relative to the clusters from ROI is close enough to 100\% so the variations related to temperature are negligible.
This cut is more likely to reject events near the edge of the detector since they have a lower number of detected photoelectrons.
The estimates of the trigger efficiency are performed with the help of the toy Monte-Carlo simulation and dedicated data from KNPP collected with gradually increasing \cevns{} trigger threshold. The stability of the count rate in ROI with described requirements is verified within OFF and ON datasets.

We apply several selection cuts on characteristics of recorded waveforms and reconstructed parameters of clusters. 
These cuts are optimized based only on the OFF data and simulated \cevns{} events by maximizing the sensitivity of the analysis to \cevns{}.
All selections as well as position and energy reconstruction procedure are applied both to simulated and measured events.

Some clusters within the defined ROI are followed by significant light emission not expected from \cevns{}. Such events can be related to enhanced local SE emission rate as well as multiple scattering of neutrons or gamma-rays. We reject events with more than 10 pulses in 5.5~$\mu$s following the identified cluster. This cut rejects about 30\% of background (OFF data) while preserving 90.5\% of livetime.

A non-negligible part of a background
is connected to coincidences of a SE electroluminescent signal with a scintillation or a short large-amplitude pulse, which can originate from the ionic afterpulses of a PMT or Cherenkov radiation in PMT glass. In such a coincidence a proper duration of a cluster is provided by SE while the most of ``energy'' is concentrated in a short time window. We suppress a contribution of these events to ROI by calculating a ratio of integral in a time window of $-$20~to~+180~ns defined relative to the onset of the largest pulse to the whole integral of a cluster. This ratio is required to be less than the conservative value of 0.4, discriminating part of the background but not affecting the acceptance of regular S2 signals. We also introduce another parameter to reject the remaining events with the Cherenkov-like pulse shape. A ratio of a mean SPE integral to a pulse amplitude is calculated for each pulse in a cluster. We restrict the lowest of these ratios to be larger than 2.0, the latter value corresponds to an amplitude of about 40~mV, i.e. about 5 times larger than an average SPE pulse. Additionally, we consider a time window of 1~$\mu$s within a cluster that has the largest fraction of integral in it. The ratio of integral within this 1~$\mu$s to the full integral is required to be lower than 0.84. Such a parameter rejects coincidences of an SE signal with low energy electroluminescence at the periphery of the detector's horizontal plane. These peripheral electroluminescent signals have a characteristic duration of 700~ns due to the peculiarity of the anode electrode design. The total efficiency of these conservative cuts relative to \cevns{} signals from ROI was estimated as 98\% based on the calculation of SPE overlap probability within a cluster.
 
\begin{figure}[hbt]
\includegraphics[width=1\linewidth]{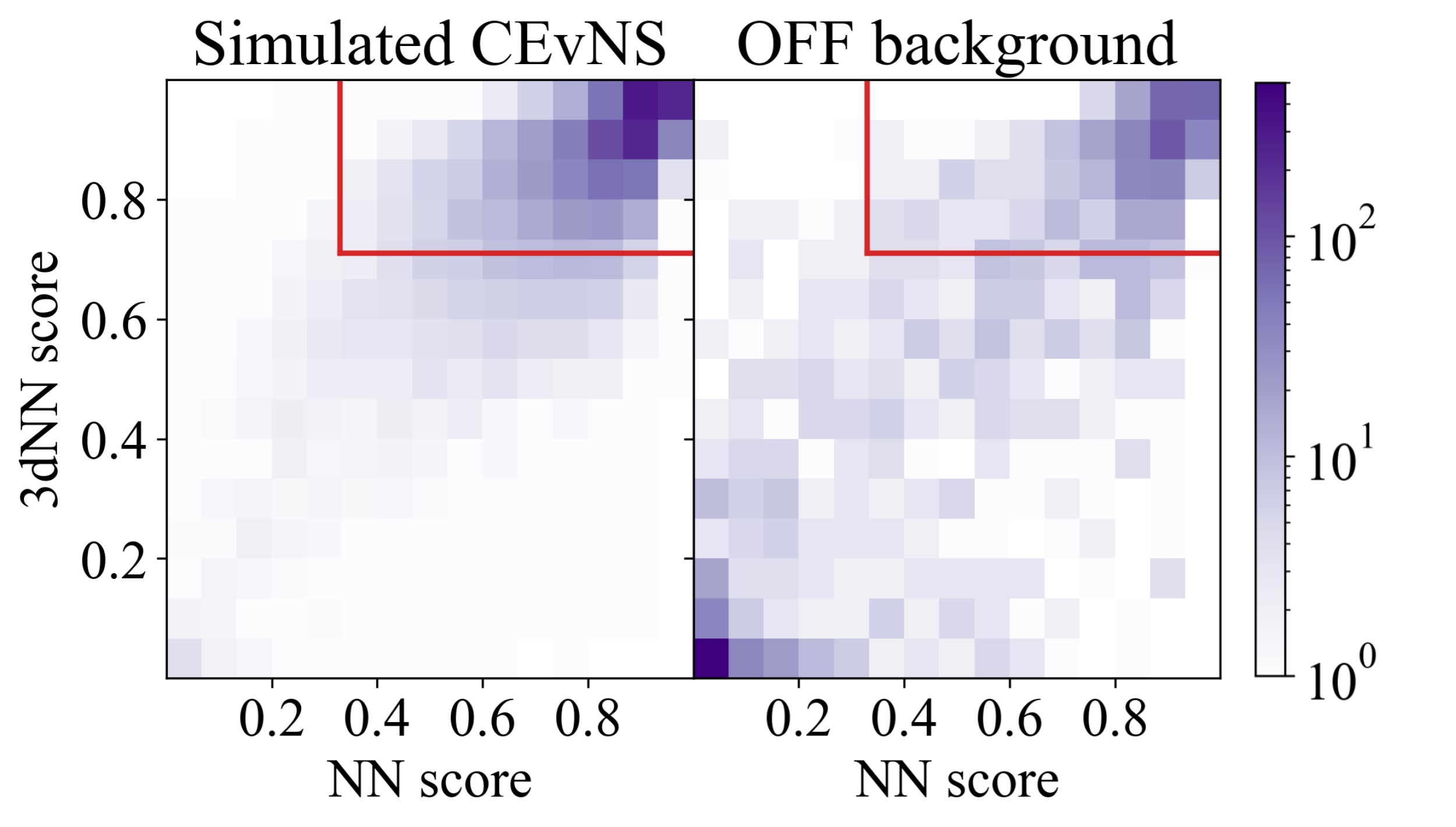}%
\caption{\label{fig:NN_comparison}
2-D distributions of the neural networks scores on the simulated \cevns{} data (left) and OFF data (right). The score means the probability to originate from the point-like source (1 -- pointlike, 0 -- not-pointlike). The red line indicates the chosen cut boundaries. Bin values for the simulated \cevns{} events are normalized to the OFF dataset size.}
\end{figure}

Finally, for each cluster, we consider the likelihood of a point-like (PL) light source in the gas gap to produce an observed distribution of light over the top PMT array. Few spontaneous SE signals can overlap producing a background event with a light distribution different from PL. To mitigate SE-coincidence background, two neural networks were developed. The first neural network (NN) uses only a light distribution over the PMT array normalized to the total amount of light registered.
It consists of 19 nodes in the input layer corresponding to 19 PMTs in the top array; 4 hidden layers with 70, 64, 72, and 44 nodes with ReLU activation function~\cite{fukushima1969visual} in each node; two batch normalization layers after the first and the last hidden layers.
The architecture of this network was optimized with KerasTuner~\cite{omalley2019kerastuner}. The second neural network (3DNN) uses 3-D (x, y, time) 10x10x20 pixels "images" of a signal.
The network includes 3 convolutional~\cite{Goodfellow-et-al-2016} layers 3x3x5 and 3 fully connected layers with ELU~\cite{ELUarticle} activation function and batch normalization after each layer.  
To predict the probability of an event being PL, the last layer in both neural networks consists of a single neuron with the sigmoid activation function. Events for training and validation datasets were simulated using the procedure described in Section~\ref{sec:simulation}. The area under receiver operating characteristic curve~\cite{rocauc} of neural networks on Monte-Carlo testing data reached 95\% with a slightly better score for 3DNN. A detailed description of NN design and training will be provided in a separate paper. We used a cut based on predictions of both networks (see figure~\ref{fig:NN_comparison}) to suppress the SE coincidence background. It can be seen that the OFF dataset contains a significant number of events with a big probability of being PL based on the NN scores. These events can be attributed to a background from interactions of ionizing radiation or to the spatial coincidence of spontaneous SE emission, e.g. originating from muon tracks~\cite{Akimov:2016rbs,Akimov:2023xsi}.

\begin{figure}[hbt]
\includegraphics[width=1\linewidth]{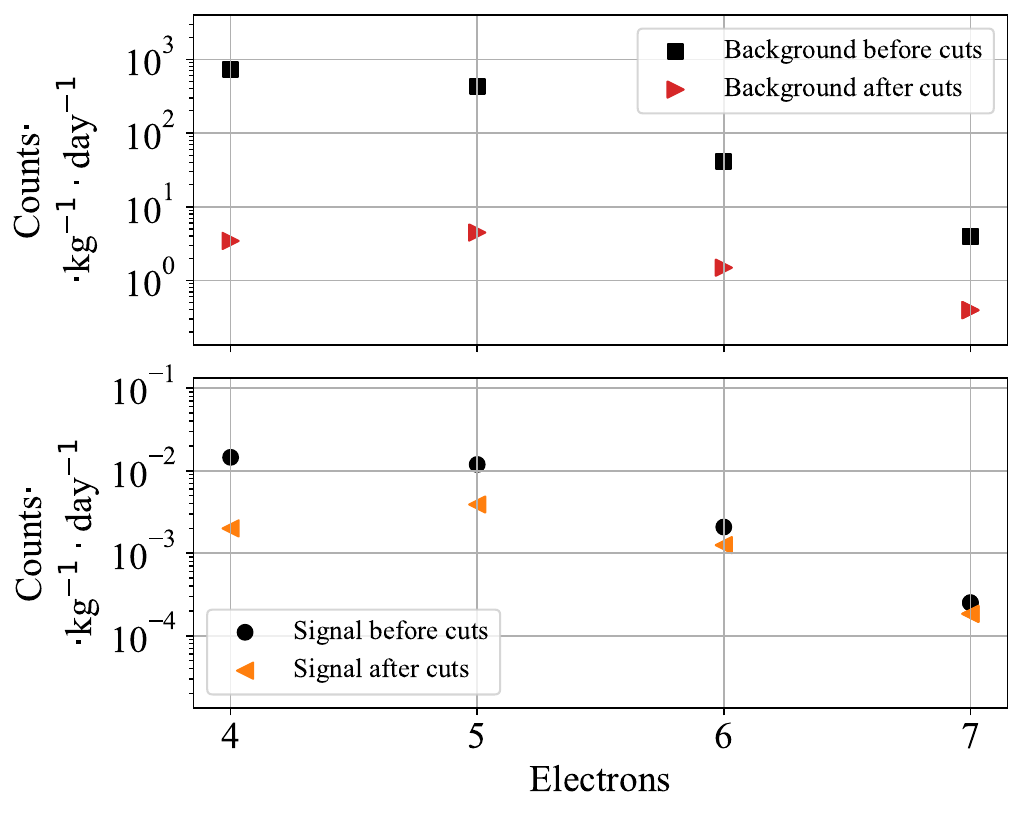}% 
\caption{\label{fig:bg_supp}Suppression of the background from reactor OFF data (top) and the \cevns{} signal (bottom) under the assumption of SM2018 spectrum model.}
\end{figure}

\begin{figure}[hbt]
\includegraphics[width=1\linewidth]{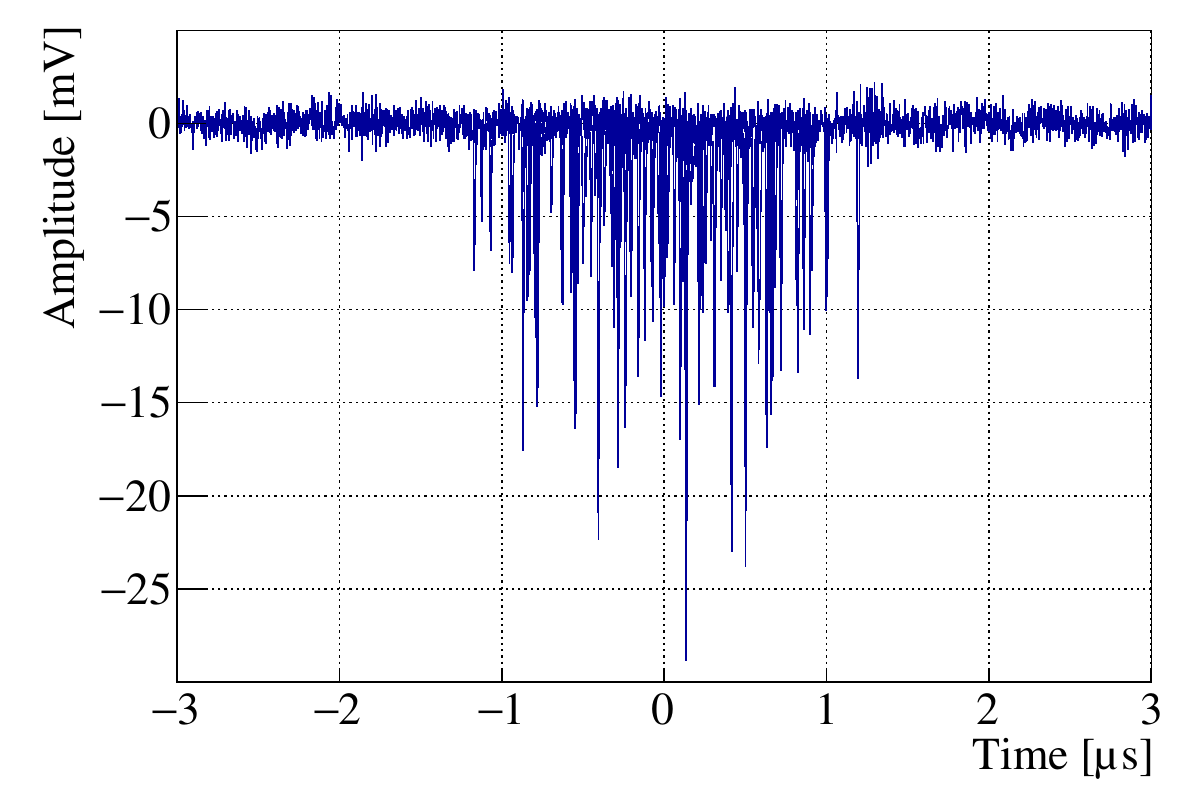}
\caption{\label{fig:event_ex}Example of an event passing all cuts successfully. Channels corresponding to PMTs of the top array are overlaid. }
\end{figure}

The resulting influence of all cuts on the background and the expected \cevns{} signal is presented in figure~\ref{fig:bg_supp}.
The background suppression is more than 99\% while the \cevns{} signal loss is 75\% in ROI.
The example of a background event passing the cuts is presented in figure~\ref{fig:event_ex}.

\section{\label{sec:sens} SENSITIVITY}

The sensitivity of RED-100 to \cevns{} is estimated based on the data acquired during the reactor OFF period and simulated signal. 
We apply the requirements described in Section~\ref{sec:processing} to the data and use parameters of clusters from the resulting selection to fill three histograms. 
The first one is for corrected energy estimates (in~PE~units). 
The second is a histogram of clusters duration connected to the vertical coordinate of an interaction through the broadening via diffusion of ionization electrons. 
The last one contains a distribution of radius squared which is calculated based on the candidates spatial position in the horizontal plane of the detector. 
We scale each of these histograms by the OFF exposition to represent the background rate in units of~counts~per~day~per~kg. In what follows we refer to them as ``scaled OFF histograms''. 
It is useful to define also three \cevns{} prediction histograms in the same parameter space and with all the selections applied (``scaled \cevns{} histograms''). Based on scaled OFF and \cevns{} histograms we evaluate the ``expected ON histograms''. The values in channels of these histograms are equal to the sum of the values in respective channels of OFF and \cevns{} spectra. The statistical uncertainties of values in expected ON histograms are recalculated to the exposition time acquired at KNPP during the reactor ON period. After that, we define three ``expected residual histograms'' by subtracting scaled OFF from expected scaled ON. By definition, the values in these residuals coincide with those from scaled \cevns{} histograms, while the statistical uncertainties are an uncorrelated sum of uncertainties from scaled OFF and expected ON. The number of counts in each channel of OFF histograms before scaling is enough to apply $\chi^2$ statistics.

We evaluate the sensitivity of the experiment to \cevns{} using the set of expected residual histograms, a so-called ``Asimov dataset''~\cite{Cowan:2010js}. The statistical analysis approach we use is based on a simultaneous fit of three residual histograms to \cevns{} expectation. The parameter of interest is the amplitude of a \cevns{} signal $A$ relative to the Standard model prediction, the only parameter varied in a fit. We consider the statistics of $\Delta\chi^2=\chi^2(A)-\chi^2(A_{best})$, where $A_{best}$ is a signal amplitude minimizing $\chi^2(A)$. For the ``Asimov'' residual $A_{best}=1$ and $\chi^2(A_{best})=0$ by definition. The 90\% confidence level~(C.L.) sensitivity, i.e. the median expected limit, can be evaluated as $A$, such that $\Delta\chi^2(A)=2.71$. The sensitivities derived for each variant of \cevns{} prediction can be found in Table~\ref{T_results} in parentheses and one of the corresponding $\Delta\chi^2$ profiles is shown in figure~\ref{fig:delta_chi2} with dashed line. It can be seen that a limit of about 60-90 times larger than Standard model \cevns{} is expected for the achieved energy threshold and exposition time. The reasons for a modest sensitivity and its significant dependence on \cevns{} signal assumptions are discussed in Section~\ref{sec:discuss}.

Statistical analysis of the residual ON$-$OFF count rate is justified if the background is stable. For the background measurements, we used several additional detectors continuously running during the whole RED-100 data-taking period. Also, regular background monitoring runs were acquired with RED-100 itself. These measurements show good background count rate stability without any significant variations with changes in the reactor operation mode. A detailed description of these measurements and obtained results can be found in the dedicated paper~\cite{Akimov:2023lsz}.

\section{\label{sec:result} RESULTS}

\begin{figure}[bht]
\includegraphics[width=1\linewidth]{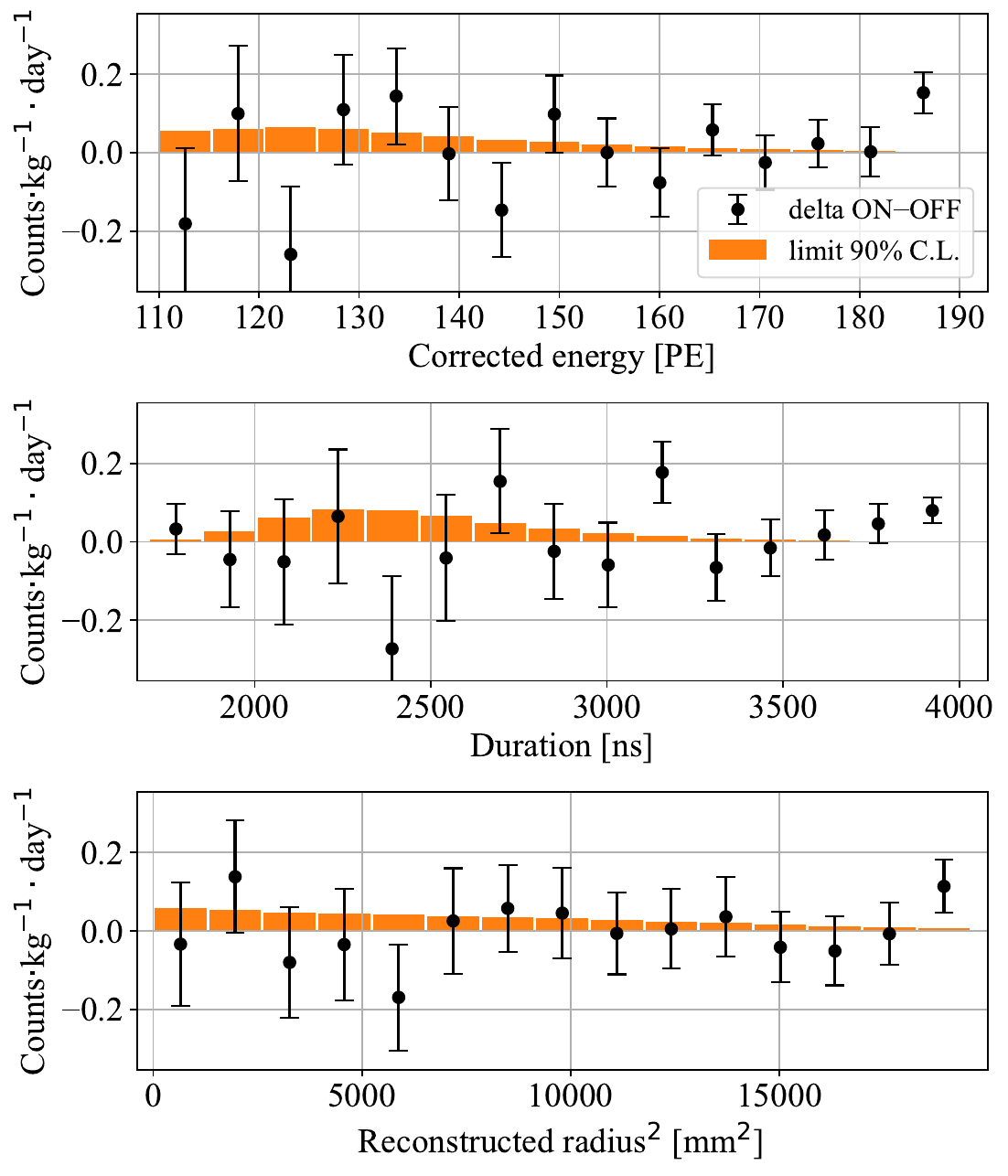}% 
\caption{\label{fig:on_off}Residual ON$-$OFF histograms for corrected energy (top), duration of a cluster (middle), and reconstructed radius squared (bottom).}
\end{figure}

\begin{figure}[hbt]
\includegraphics[width=1\linewidth]{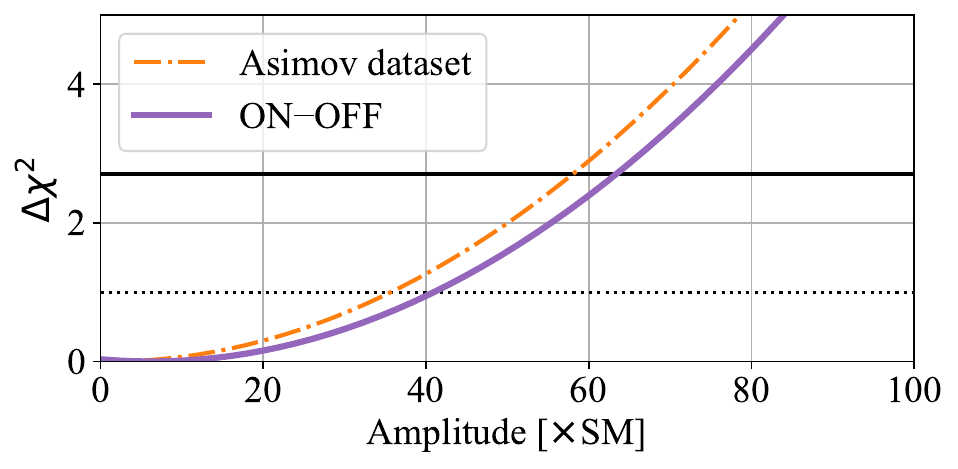}% 
\caption{\label{fig:delta_chi2} Profiles of $\Delta\chi^2$ for the sensitivity studies and constraints of \cevns{} amplitude under assumption of SM2018.}
\end{figure}

To evaluate experimental limits on \cevns{} amplitude we repeat the analysis from Section~\ref{sec:sens}, but substitute expected count rate histograms and uncertainties with real ones based on the reactor ON data. Corresponding residual ON$-$OFF spectra are presented in figure~\ref{fig:on_off}. When fit to the \cevns{} predictions they allow to produce $\Delta\chi^2$ profiles like the one shown in figure~\ref{fig:delta_chi2} and evaluate the experimental limits summarized in Table~\ref{T_results}. It can be seen that the best fit amplitude does not contradict the Standard model \cevns{} prediction within the statistical uncertainty. The evaluated upper limits are slightly larger than the sensitivities calculated for the corresponding variants of a \cevns{} prediction.

\begin{table}[hbt]
\caption{\label{T_results} Upper limits (sensitivity) on \cevns{} amplitude depending on the models of a primary antineutrino spectrum.}
\begin{center}
\begin{tabular}{ |p{1.7cm}|p{1.7cm}|p{1.7cm}|p{1.7cm}| } 
 \hline
 \multicolumn{4}{|c|}{Limit (Sensitivity) at 90\%~C.L., $\times$SM} \\
 \hline
  \centering SM 2018 & \centering KI & \centering DB & \centering INR \tabularnewline
 \hline
 \centering  63 (58) & \centering 94 (90) & \centering 61 (56) & \centering 70 (64) \tabularnewline
 \hline
\end{tabular}
\end{center}
\end{table}

Apart from the systematic uncertainty associated with the reactor antineutrino energy spectra, there are two more effects significantly affecting the strength of the evaluated limit. We illustrate the impact of these effects under the assumption of the SM2018 antineutrino spectrum. The first one is the uncertainty of the nuclear recoil charge yield in xenon estimated within the NEST framework. The change of the mean NEST charge yield to the lower (upper) edge of the corresponding uncertainty band (see~Fig.3,~bottom in ref.~\cite{szydagis2023review}) shifts the limit to 135$\times$SM (27$\times$SM). More low-energy nuclear recoil data for the NEST input are required to suppress this effect. Another source of uncertainty is the EEE value of 32.8$\pm$2.8\%. Variation of EEE down~(up) within its standard deviation changes the limit to 78$\times$SM (43$\times$SM) for the default nuclear recoil charge yield. The EEE value uncertainty evaluated for the RED-100 exposition at KNPP is dominated by the accuracy of the average energy required to produce an excitation quantum in liquid xenon $W=13.8\pm0.9$~eV~(see refs.~\cite{CalibrationRED100,Doke_2002}).

\section{\label{sec:discuss} DISCUSSION}

The modest sensitivity of the first run of RED-100 is caused by a combination of factors: relatively high energy threshold, higher than expected background rate, and moderate exposition time.
The energy threshold of statistical analysis presented in this work is about 110 PE (4 ionization electrons). 
Given the achieved electron extraction efficiency of $32.8\pm2.8$~\% and the ionization yield of xenon from NEST~\cite{szydagis2023review} this threshold corresponds to about 0.2~keV of electron recoil and 2~keV of nuclear recoil equivalent. 
The maximum energy of a xenon recoil from \cevns{} of a 8~MeV antineutrino is about 1~keV. 
It means that only nuclear recoils from the highest-energy antineutrinos, producing a signal smeared by the fluctuations of ionization yield, extraction to the gas gap, and light collection end up within the RED-100 analysis ROI. 
This consideration explains a significant dependence of the evaluated sensitivity estimates on the model of the reactor antineutrino spectrum (see Table~\ref{T_results}). The expected energy threshold was not achieved due to the trigger efficiency instability associated with significant temperature variations at the site of the experiment.

%bg v2
The background rate observed in ROI at KNPP is significantly higher than expected~\cite{RED-100:2019rpf}. Our simulations allow us to conclude that both ambient and cosmogenic neutrons as well as gamma rays cause a count rate much lower than that observed~\cite{Akimov:2023lsz}. 
Another expected source of background is associated with spontaneous single electron emission. This process significantly affects the operation of two-phase xenon detectors at a shallow overburden. The SE rate observed at KNPP is reduced by about an order of magnitude compared to the previous measurements at MEPhI, down to 25~kHz~\cite{Akimov:2023lsz}.
The laboratory tests of the RED-1 prototype and the RED-100 detector demonstrated that the SE events flow is not Poissonian and consequent single electron signals are correlated in space~\cite{Akimov:2016rbs,Akimov:2023xsi}.
The presence of such correlations drastically undermines the PL cut efficiency.
Correlated SE emission may be possible explanation of the observed background excess. Additional hints to this hypothesis include decreasing of the background rate with energy (lower probability of more electrons to overlap) and increasing for a larger duration at a given energy (more probable overlap within a longer time window).

The moderate time of the detector exposition at KNPP is connected to technical issues on site. %
In order to estimate a potential of \cevns{} observation we extrapolate the sensitivity of RED-100 to an astronomical year of operation at KNPP: a month of reactor outage and eleven months of reactor operation. This extrapolation suggests an expected 90\%~C.L. limit of about 15-20 times above the Standard Model prediction, still insufficient to observe \cevns{}.

Despite the ability of RED-100 to detect single ionization electrons its sensitivity to \cevns{} is limited by a combination of a background and low energy of xenon nuclear recoils. We consider the change of the active medium from xenon to argon, which allows for larger energy of nuclear recoils and larger electron extraction efficiency for the same electric field strength~\cite{Gouschin1978}. The nuclear recoil charge yield is comparable for xenon and argon in reactor \cevns{} ROI~\cite{NEST_BM}. The drawbacks of using argon include lower total \cevns{} cross-section, $^{39}$Ar-related background, and lower light yield of electroluminescence in combination with a challenge of 128~nm light detection.
The preliminary estimates of \cevns{} count rate in the argon-filled RED-100 can be found in ref.~\cite{physics5020034}. These estimates suggest that the neutrino signal can be observed over the $^{39}$Ar beta-background in the energy deposition range below five ionization electrons. It is not clear if the rate of spontaneous single electron signals in argon is lower or larger than in xenon. The upcoming laboratory tests of RED-100 with argon are to answer this question and show if the change of the medium is beneficial for the detector's sensitivity to reactor \cevns{}.

In the course of evaluation of the \cevns{} count rate we encountered an aspect of calculation we would like to highlight. It is a simulation of ionization yield fluctuations of nuclear recoils. While the results presented in this work are based on the fluctuations model from NESTv2~\cite{szydagis2023review}, there are other approaches to taking this effect into account, e.g. considered in refs.~\cite{PhysRevLett.116.161301,Aprile_2016,James_2022}. The difference between models is due to the application of particular fluctuation statistics at the steps of generation of total quanta (light and charge), number of ions, recombination, and quenching. Verification of these models is complicated by the scarcity of the low energy data both for xenon and argon. Though our tests show significant dependence of the \cevns{} count rate above the detector threshold on the ionization yield fluctuation model, the NESTv2 approach gives the most conservative result. We note that the dependence of the result on the fluctuation model wanes with the reduction of the energy threshold and improvement of the detector resolution.

\section{\label{sec:conclusion}Conclusion}

The first run of RED-100 at Kalinin Nuclear Power Plant demonstrates the feasibility of a 100~kg-scale two-phase noble gas detector operation at NPP with a low threshold of about 4 ionization electrons. The data analysis shows no statistically significant difference between 331 (192 in FV) kg$\cdot$days of reactor ON and 106 (61 in FV) kg$\cdot$days of reactor OFF. We obtain the first constraint on the coherent scattering of reactor antineutrinos off xenon nuclei of about 60-90 times larger than the Standard Model prediction, though dependent on the model of the primary antineutrino energy spectrum and ionization yield fluctuations.
The performance of RED-100 during the first run at KNPP is significantly affected by the single electron background associated with high energy depositions. A~detailed study of spatial and time correlations between SE signals would greatly benefit understanding of the background in the range energy deposition range of a few ionization electrons. 
The sensitivity of the RED-100 experiment extrapolated to an astronomical year of data taking at KNPP is comparable to the first phases of other reactor experiments, although insufficient for \cevns{} detection. The change of the sensitive medium from xenon to argon is considered to facilitate observation of \cevns{} at a nuclear reactor.

\begin{acknowledgments}

The authors express their gratitude to the State Atomic Energy Corporation Rosatom (ROSATOM) and the Rosenergoatom Joint-Stock Company for administrative support of the RED-100 project, the JSC Science and Innovations (Scientific Division of the ROSATOM) for the financial support under contract No.313/1679-D dated September 16, 2019, the Russian Science Foundation for the financial support under contract No.22-12-00082 dated May 13, 2022, the administrations of the National Research Nuclear University MEPhI (MEPhI Program Priority 2030), the National Research Center “Kurchatov Institute”, the Institute of Nuclear Physics named after G.I. Budker SB RAS. 
The work was funded by the Ministry of Science and Higher Education of the Russian Federation, Project “New Phenomena in Particle Physics and the Early Universe” FSWU-2023-0073. The authors are grateful to the director of the Institute of Industrial Nuclear Technologies of MEPhI (IINT~MEPhI) - Eduard Glagovsky and the management of the Institute of Nuclear Physics and Technologies of MEPhI (INPT~MEPhI), for their support.
Our work was performed using resources of NRNU MEPhI high-performance computing center. The authors are grateful to the staff of the Kalinin NPP for their comprehensive assistance in conducting the RED-100 experiment, as well as the scientists from DANSS, $\nu$GeN, and iDREAM experiments at the Kalinin NPP, for assistance in organizing the measurements. We would like to add a special thanks to Olga Zeldovic (Kurchatov Institute) for productive discussions of data analysis and signal selection. We also would like to express our gratitude to 
Matthew Szydagis~(SUNY Albany) for fruitful discussions about nuclear recoil charge and light yield models.
We are thankful to Valery Sinev~(INR~RAS, MEPhI) and Nataliya Skrobova~(LPI~RAS) for exchanges regarding the reactor antineutrino energy spectra.
Dmitry Rudik and Olga Razuvaeva express their gratitude to Andrey Voynov~(Google) and Leonid Gremyachikh~(HSE) for the fruitful discussions about possible neural network implementations for background suppression.
\end{acknowledgments}

\nocite{*}

\bibliography{apssamp}% Produces the bibliography via BibTeX.

\end{document}